# Magneto-electrical subbands of freely suspended quantum point contacts

C. Rossler,[1] M. Herz,[1] M. Bichler,[2] and S. Ludwig[1]

[1]Fakultät für Physik and Center for NanoScience, Ludwig-Maximilians-Universität, Geschwister-Scholl-Platz 1, D-80539 München, Germany.

[2]Walter Schottky Institut and Physik Department, Technische Universität München, Am Coulombwall 3, D-85748 Garching, Germany.

We present a versatile design of freely suspended quantum point contacts with particular large one-dimensional subband quantization energies of up to $\Delta\varepsilon \approx 10\text{ meV}$. The nanoscale bridges embedding a two-dimensional electron system are fabricated from AlGaAs/GaAs heterostructures by electron-beam lithography and etching techniques. Narrow constrictions define quantum point contacts that are capacitively controlled via local in-plane side gates. Employing transport spectroscopy, we investigate the transition from electrostatic subbands to Landau-quantization in a perpendicular magnetic field. The large subband quantization energies allow us to utilize a wide magnetic field range and thereby observe a large exchange splitted spin-gap of the two lowest Landau-levels.

Since the first experimental realization [1,2], quantum point contacts (QPCs) prove to be versatile devices, both regarding fundamental physics and applications. Although the basic theory of a one-dimensional electronic system is well established [3], there are still open questions regarding the interactions between charge carriers, e.g. the origin of the 0.7 anomaly is still a subject of discussion.[4,5] Recent experiments demonstrate that driven QPCs act as emitters of phonons.[6,7] Freely suspended QPCs, as presented here, will allow further investigations of the electron-phonon interaction and might contribute to the ongoing debate in distinguishing phononic interaction from Coulomb-interaction in nanostructures.[8,9] Moreover, due to their electrical decoupling from the substrate, freely suspended structures prove to be well suited for optoelectronic devices [10,11]. The steep *I-V*-characteristic of QPCs combined with their weak Coulomb-screening allows to utilize them for detecting small fractions of the elementary charge in nearby quantum dots, making QPCs a tool of choice for the read-out of quantum dot qubits.[12,13] To avoid inter-mode-scattering, a large energy spacing between the one-dimensional (1D) subbands is favorable. The highest reported subband-spacings in GaAs of $\Delta\varepsilon \approx 10...20~\text{meV}$ have been accomplished in shallow-etched QPCs [14]. In our QPCs we find subband-spacings of up to $\Delta\varepsilon \approx 10~\text{meV}$. Here, we investigate the transition from the one-dimensional subband-quantization of a QPC at zero magnetic field to Landau-quantization at finite magnetic fields $B_\perp$ oriented perpendicular to the plane of the two-dimensional electron system (2DES). In the past, the transition was observed at temperatures $T < 1~\text{K}$ and within a relatively narrow magnetic field range at $B_\perp < 2~\text{T}$.[15-17] The advantage of our large subband-spacings is that this transition is shifted to larger magnetic fields allowing a better resolution even in a measurement at a higher temperature of $T = 4.2\,\text{K}$.

Our fabrication starts with a modulation doped GaAs/AlGaAs heterostructure consisting of a $130\,\text{nm}$ thick active layer on top of $400\,\text{nm}$ of an $Al_{0.8}Ga_{0.2}As$ sacrificial layer [18,19]. The active layer incorporates a $25\,\text{nm}$ thick GaAs quantum well where the 2DES resides, located approximately $40\,\text{nm}$ beneath the surface of the heterostructure. The initial electron sheet density $n_S \approx 5.5 \times 10^{11}\,\text{cm}^{-2}$ and mobility $\mu \approx 7.8 \times 10^5\,\text{V}^{-1}\text{s}^{-1}$ of the 2DES are both reduced by a factor of typically two to five, caused by etching damage during processing. First, ohmic contacts are defined at non-suspended areas using annealed AuGeNi pads to allow electric connection of the 2DES. Then, a $60\,\text{nm}$ thick layer of nickel is deposited by means of optical lithography (outer area) and electron-beam-lithography (center region) to protect specific areas of the active layer. Anisotropic reactive ion etching using $SiCl_4$ removes the uncovered top layers of the heterostructure. The nickel layer is then removed by $FeCl_3$. Finally, isotropic wet-etching with 1% hydrofluoric acid selectively removes the $400\,\text{nm}$ thick sacrificial layer in the vicinity of the etched trenches. The final result are free-standing beams containing a 2DES. The investigated sample is presented in a scanning-electron-micrograph in Figure 1(a). The GaAs substrate appears dark gray, whereas completely suspended areas of the active layer are light gray. Note that the active layer appears darker wherever the distance to a trench-edge is larger than approximately $2\,\mu\text{m}$. Here, the 2DES is no longer underetched. This scattering-effect allows us to conclude that the (lights gray) center beam connecting source and drain is indeed freely suspended. The dimensions of the bridge are $4000\,\text{nm} \times 600\,\text{nm} \times 130\,\text{nm}$ (length x width x thickness). It contains a central constriction reduced to a width of $300\,\text{nm}$, defining the QPC, which can be capacitively controlled by applying voltages to adjacent 2DES areas (sidegates G1 and G2) [20]. Figure 1(b) shows the same structure under a tilt angle of 75 degrees. Having fabricated several similar samples, we observe the trend that constrictions smaller than about $250\,\text{nm}$ are electrically isolating. This indicates a lateral depletion length $l_{\text{DEP}} \approx 125\,\text{nm}$ of the 2DEG, being in good agreement with comparable non-underetched structures [21].

After initial illumination at $T = 4.2\,\text{K}$ using an LED ($\lambda = 950\,\text{nm}$, $P \approx 1\,\text{mW}$, $t \approx 30\,\text{s}$), the two-terminal differential conductance of the device is measured as a function of the voltage applied to both gates G1 and G2 (lock-in frequency $f = 84\,\text{Hz}$, modulation amplitude $\delta V_{\text{SD}} = 20\,\mu\text{V}$). The resulting linear response pinch-off curve shown in Figure 1(c) exhibits the characteristic conductance quantization of a QPC. A serial lead resistance of order $R_L \sim 10\,\text{k}\Omega$ is already subtracted ($g^{-1} = g^{-1}_{\text{MEASURED}} - R_L$) where the exact value is chosen to assure conductance plateaus at the theoretical expected values of $g = n \cdot 2\,\text{e}^2/\text{h}$ ($n = 1,2,3,4$). The rather large value for $R_L$ reflects the large resistance of the suspended bridge while the ohmic contacts have resistances of $R_{\text{CONT}} < 1\,\text{k}\Omega$. The differential conductance is plotted in Figure 2 as a function of the gate voltages $V_{G1} = V_{G2}$ and the source-drain bias $V_{\text{SD}}$ [15]. The first three diamond shaped conductance plateaus with $n = 1,2,3$ are highlighted by dashed lines. The vertical width of these diamonds ranging from $-20\,\text{mV} \leq V_{\text{SD}}^{\text{DIAMOND}} \leq 20\,\text{mV}$ (arrows in Fig. 2) is directly related to the 1D subband spacings via $\Delta\varepsilon_n = V_{SD}^{DIAMOND} / (1 + R_L \cdot 2ne^2/h)$. Our first two 1D-subbands are spaced by $\Delta\varepsilon_{12} \approx 10.8\,\text{meV}$, while $\Delta\varepsilon_{23} \approx 7.4\,\text{meV}$ and $\Delta\varepsilon_{34} \approx 5.6\,\text{meV}$. In contrast to the integer plateaus, the so-called half plateaus (labeled '0.5' to '3.5' in Fig. 2) are obscured by strong noise, as they combine a non-linear transport regime with incomplete transmission.

Figure 3(a) displays the linear response differential conductance as a function of the gate voltage $V_{G1}$ and a magnetic field $B_\perp$ perpendicular to the plane of the 2DES. In order to compensate for the strong modulation of the conductance caused by Shubnikov-de-Haas oscillations in the leads, $R_L$ has been adjusted separately for each trace of constant $B_\perp$. For comparison the same data are shown in Figure 3(b) assuming $R_L = 12\,\text{k}\Omega$ independently of $B_\perp$. The onsets of integer conductance plateaus (bottom of 1D-subband at Fermi energy) with $g = n \cdot 2\,e^2/h$ and $n = 1,2,3,4$ are marked by open circles in Figure 3(a). The transition from 1D-subbands independent of $B_\perp$ to Landau-levels of Energy $E_n = (n - 1/2) \cdot \hbar\omega_C$, with the cyclotron frequency $\omega_C = eB_\perp / m_e^*$, is clearly visible. Here, $n = 1,2,...$ stands for the spin-degenerate $n$-th Landau-level and the conductance plateaus are described by spin-degenerate incompressible strips with integer filling factors $\nu = 2n$. In addition, for $B_\perp \geq 5\,\text{T}$ we observe two plateaus with filling factors $\nu = 1$ and $\nu = 3$, reflecting a lifted spin-degeneracy (onset marked in Fig. 3(a) by closed circles). The transition from the 1D-subbands of a QPC at $B_\perp = 0\,\text{T}$ to spin split Landau-levels can be approximated (assuming a symmetric spin splitting) by $E_n = V_{G1} \cdot \alpha(V_{G1}, B_\perp) = E_0(V_{G1}, B_\perp) + [n - 1/2]\sqrt{\Delta\varepsilon_n^2(B_\perp = 0, V_{G1}) + [\hbar\omega_C]^2} \pm 1/2 \cdot g^* \mu_B B_\perp$ [16,17] with an effective g-factor $g^*$, the Bohr magneton $\mu_B$ and the bottom of the 1D-confinement potential $E_0$ (taking the Fermi energy $E_F \equiv 0$ as reference). In our case, both $E_0$ and the conversion factor $\alpha$ (correlating the $V_{G1}$-axis in Figure 3 to energy) are expected to depend on $V_{G1}$ because the screening properties of the electron system within the narrow bridge depend on the gate voltage.[22,23] Moreover, $\alpha$ is expected to depend on the gate voltage because the 1D density of states of the QPC is energy-dependent. In our samples however, the latter dependency is found to be relatively weak due to disorder-broadening. Hence, we use for each subband a constant effective conversion factor as an approximation. Note that a careful analysis of the involved energy scales suggests that our QPC is best described by a one-dimensional electron density of states even for our largest $B_\perp$ applied. This is crucial for the applicability of the above model to the data presented in Figure 3.

The solid lines in Figure 3(a) express this model assuming $g^* = 0$, hence cutting right through the $\nu = 1$ and $\nu = 3$ plateaus. Starting from an approximation $\Delta\varepsilon(B_\perp = 0, V_{G1}) = 8.0 \text{ meV} - 0.35 \cdot V_{G1}$ (taken from a direct measurement of the 1D subband spacings – compare Figure 2) we find the following fit parameters for the onsets of the plateaus with $g = n \cdot 2 \text{ e}^2/h$: Sorted by index $n = \{1,2,3,4\}$ the conversion factor is $\alpha = \{1.19, 0.83, 0.77, 0.77\} \text{ meV/V}$ and the bottom of the QPC´s confinement potential is $E_0 = \{-4.6, -10.4, -12.9, -13.1\} \text{ meV}$, which is consistent with the aforementioned considerations.[22] Compared to an analysis of the diamonds in Figure 2, we find $\alpha$ to be reduced by approximately a factor of 2, which we explain as follows. As the magnetic field modulates the density of states and, if strong enough, causes localization of the electronic wave functions, we expect $\alpha$ and $E_0$ to also depend on $B_\perp$. The observed reduction of $\alpha$ suggests stronger screening at larger magnetic fields.

The onset of the $\nu=1$ and $\nu=2$ ($n=1$) plateaus are visible as local extrema of $dg/dV_{G1}$ (numerically differentiated) displayed in Figure 3(c). The apparent splitting (see also Fig. 3a) is an approximately linear function of $B_\perp$. Within our approximation, we find a large effective g-factor of $g^* \approx 6$ at $B_\perp = 10\,\mathrm{T}$ (arrows in Fig. 3c) which we attribute to the exchange interaction between the spins.[24-27] Interestingly, in our case the magnetic field dependent effective g-factor of the second Landau-level is comparable to $g^*(B_\perp)$ for $n=1$. Thus our result indicates that the expected screening of the exchange coupling is reduced in a 1D-QPC.

In summary, we present transport experiments on freely suspended quantum point contacts. We find 1D-subband-spacings of these devices of up to $\Delta\varepsilon \approx 10\,\mathrm{meV}$, being amongst the highest values reported in QPCs so far. Investigating the transition of 1D-subbands of a QPC to Landau-levels in a magnetic field, we find large exchange-enhanced spin gaps of similar size in the lowest two Landau-levels.

The authors thank J. P. Kotthaus, V. T. Dolgopolov and V. S. Khrapay for fruitful discussions. We gratefully acknowledge financial support from BMBF via nanoQUIT, the DFG via DIP and the German excellence initiative via the "Nanosystems Initiative Munich (NIM)" and LMUinnovativ.

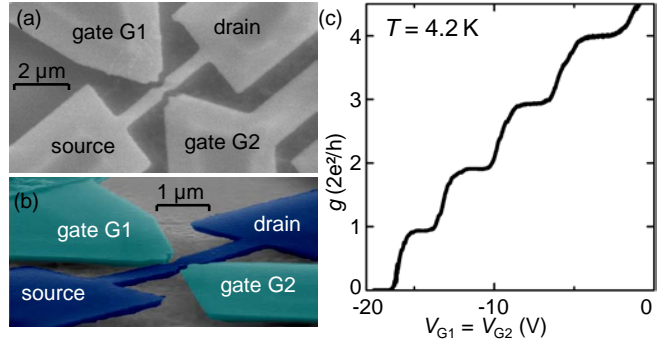

FIG. 1. (color online) (a) Scanning electron micrograph of the device. The GaAs substrate appears dark, the active layer resides on a socket which is edgewise undercut (pale stripes). The areas labeled 'source' and 'drain' are connected via a fully suspended beam with dimensions of $4000\,\text{nm} \times 600\,\text{nm} \times 130\,\text{nm}$ (length x width x thickness). One etched $300\,\text{nm}$ wide constriction defines a QPC. Two adjacent areas containing 2DESs are employed as side gates G1 and G2. (b) Side view of the device under a tilt angle of 75 degree. (c) Two-terminal differential conductance g in units of $2e^2/h$ as a function of the gate voltages $V_{G1} = V_{G2}$. A lead resistance of $R_L \approx 10\,\text{k}\Omega$ has been subtracted ($T = 4.2\,\text{K}$, $V_{SD} = 0\,\text{mV}$).

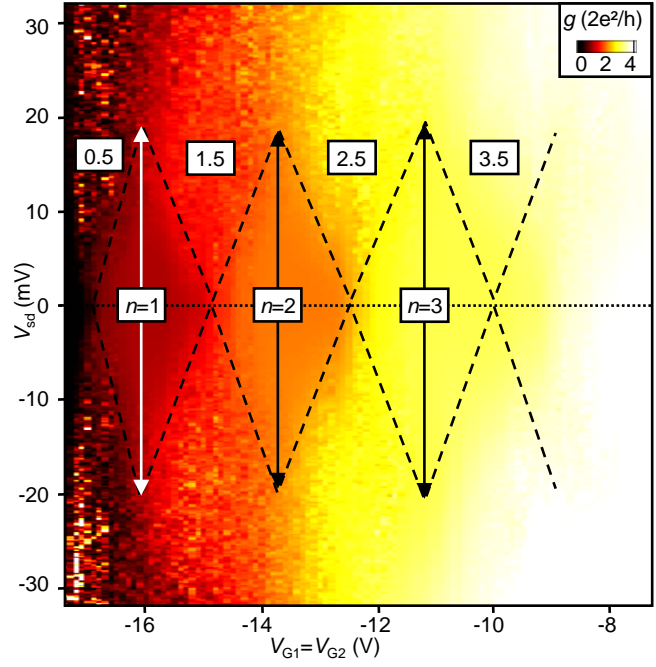

FIG. 2. (color online) Differential conductance g in units of $2e^2/h$ as a function of the source-drain-bias $V_{SD}$ and gate voltage $V_{G1} = V_{G2}$ (lead resistance of $R_L \approx 10\,k\Omega$ subtracted, $T = 4.2\,K$). Dashed lines frame the diamond shaped integer conductance plateaus of the first three subbands ($g = n \times 2e^2/h$, $n = 1, 2, 3$).

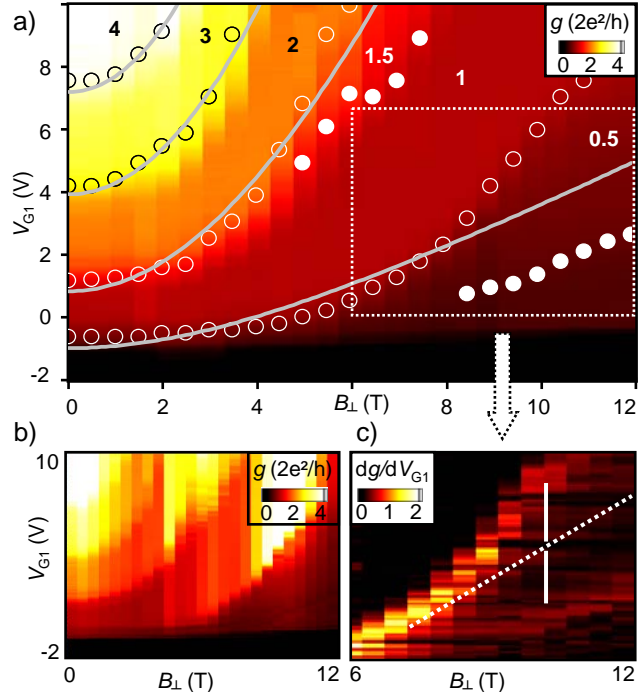

FIG. 3. (color online) (a) Differential conductance g in units of $2e^2/h$ as a function of the gate voltage $V_{G1}$ and a perpendicular magnetic field $B_\perp$ ($T = 4.2\,K$). The lead resistance is fitted separately for each trace of constant $B_\perp$ to ensure conductance plateaus with $g = n \times 2e^2/h$ (compare numbers in figure $n = 1,2,3,4$, while numbers 0.5 and 1.5 refer to odd filling factors $\nu = 1$ and $\nu = 3$). Open (closed) circles mark the onsets of conductance plateaus with integer (half integer) subband index. Solid lines are fit-curves. (b) Raw data, after subtraction of a constant lead resistance $R_L = 12\,k\Omega$. (c) Transconductance $dg/dV_{G1}$ obtained by numerically differentiating the framed data (white rectangle in (a)). The onset of plateaus (conductance steps) appear as local maxima. The first Landau level with energy $E_0 = 0.5 \times \hbar\omega_C$ (dashed white line) is assumed to reside in the center of the spin-gap (vertical arrows).